\documentclass[aip,amsmath,preprint,onecolumn,14pt,groupedaddress]{revtex4-1}

\usepackage[english]{babel}
\usepackage{graphicx}
\usepackage{color}
\usepackage{hyperref}
\usepackage{bm}
\usepackage{relsize}
\usepackage{amsmath}

\DeclareGraphicsExtensions{.eps,.pdf,.png,.jpg}

\hypersetup{colorlinks=false}

\begin{document}

\title{Squeezed State of an Electron Cloud as a ``Quasi-Neutral'' One-Component  Plasma }

\author{Y. Bliokh}
\email{bliokh@physics.technion.ac.il}
\affiliation{Physics Department, Technion, Israel Institute of Technology, Haifa 320003, Israel}
\author{J.G. Leopold}
\affiliation{Physics Department, Technion, Israel Institute of Technology, Haifa 320003, Israel}
\author{Ya.E. Krasik}
\affiliation{Physics Department, Technion, Israel Institute of Technology, Haifa 320003, Israel}

\begin{abstract}
 We present a one-dimensional model which gives a novel physical interpretation to the specific state of an ensemble of electrons continuously injected into an electrostatic potential well immersed in a strong applied magnetic field preventing radial expansion. When the space-charge field of the electrons accumulated in the potential well compensates the external electrostatic field, a force-free steady-state of the electron cloud forms. This state of equilibrium is known in another context as a squeezed state of an electron beam. It is shown that the spatial distribution of the electron number density in this steady-state correlates with the shape of the potential well. Perturbations of the steady-state propagate along the electron cloud in the form of Trivelpiece-Gould modes.
	
\end{abstract}

\maketitle

\section{Introduction}

The term ``squeezed state'' of an electron beam was introduced in Ref. \cite{Ignatov-1994}.  The authors of this article numerically and analytically studied the injection of a magnetized electron beam into a conducting tube consisting of two sections of increasing radii. This configuration is in frequent use in experiments with virtual cathodes and microwave oscillation sources such as vircators (see, e.g., \cite{VC} and references therein). It was shown that, under certain conditions, a virtual cathode can be tailored to appear in the larger radius section, which then moves towards the electron source placed in the smaller radius section, leaving behind it a dense cloud of low-energy electrons in a specific state, which the authors named a ``squeezed beam''. This is a state characterized by strong contraction of the electron phase space associated with the longitudinal degree of freedom.

A similar phenomenon, that is, strong contraction of the electron phase space was noticed in Ref. \cite{Bettega-2007}, where the electron injection in the Malmberg-Penning trap \cite{MP-trap,MP-trap-2} was studied numerically and experimentally. This contracted state, which the authors named  ``pure electron plasma'', has similar characteristics to the ``squeezed beam'' state, namely, low electron energies and equal, or almost equal, forward and backward electron fluxes. Because of this, the term ``squeezed state'' is more adequate than ``squeezed beam'' and will be used below.

Recently, the squeezed state of the electrons attracted particular interest in connection with its possible application in microwave electronics as a source of electrons instead of traditional cathodes \cite{Shamiloglu, Fuks, Siman-Tov, Leopold}. The squeezed state of the electrons is created between two components of a split cathode, that is, an emitter and a reflector connected by an axial conductor \cite{Siman-Tov}. The electrons accumulated in the space between the emitter and the reflector can be used as the electron source in a situation where a magnetron anode block encircles the split cathode. For this application the electron source is only partially squeezed because of the presence of other forces (radial and azimuthal) which cause electrons to drift to the anode. There are other proposals to use the squeezed electron state for microwave generation, \cite{Egorov, Dubinov-2016,Dubinov-2017}, gas ionization \cite{Dubinov-IEEE}, etc.

In this present article we study the formation of the squeezed state for electrons injected into a 1D electrostatic potential well immersed in a strong applied magnetic field. This configuration is similar to the Malmberg-Penning trap which is used for the long-time confinement of charged particles, electrons in particular. Despite this similarity, there is an important qualitative distinction between the properties of the ensembles of the charged particles in the Malmberg-Penning trap and in the squeezed state.

The formation of the equilibrium electron cloud in the Malmberg-Penning trap can be considered as a two-step process. First, electrons are injected through an ``open'' end of the potential well into the trap. Then, the potential at this end is varied to the value required for the electrons to be trapped in the well. The trapped electrons form a  \textit{closed (autonomous)} dynamical system, which evolves to thermodynamic equilibrium  \textit{without interaction with the environment} (radiative cooling is not important at this stage). The equilibrium electron cloud forms in this second step. The parameters of this steady state are defined by integrals of motion   (detailed description and references one can find in Ref. \cite{Dubin-1999}).

The squeezed state is formed in a different way. Electron injection is continuous, the environment (the electron source and the beam) and the electron cloud inside the potential well exchange particles the entire time. Thus, the squeezed state is an  \textit{open} system unlike the {\textit closed} Malmberg-Penning trap. 

In this article, we propose a new theory for the formation of the squeezed state, which allows for the determination of parameters, such as the longitudinal distribution of the electron number density, the accumulated charge, etc. To be specific, we study the evolution of electrons continuously injected into an {\textit open} Malmberg-Penning trap, a situation allowing a rather simple and illustrative interpretation.

The remainder of this article is organized as follows.  Section II expands on the description of the squeezed state.  Section III present the results of numerical simulations.  Section IV describes the squeezed state as a one-component plasma.  The conclusions are presented in Section V.

\section{The squeezed state}  

\begin{figure}[tbh]
	\centering \scalebox{0.1}{\includegraphics{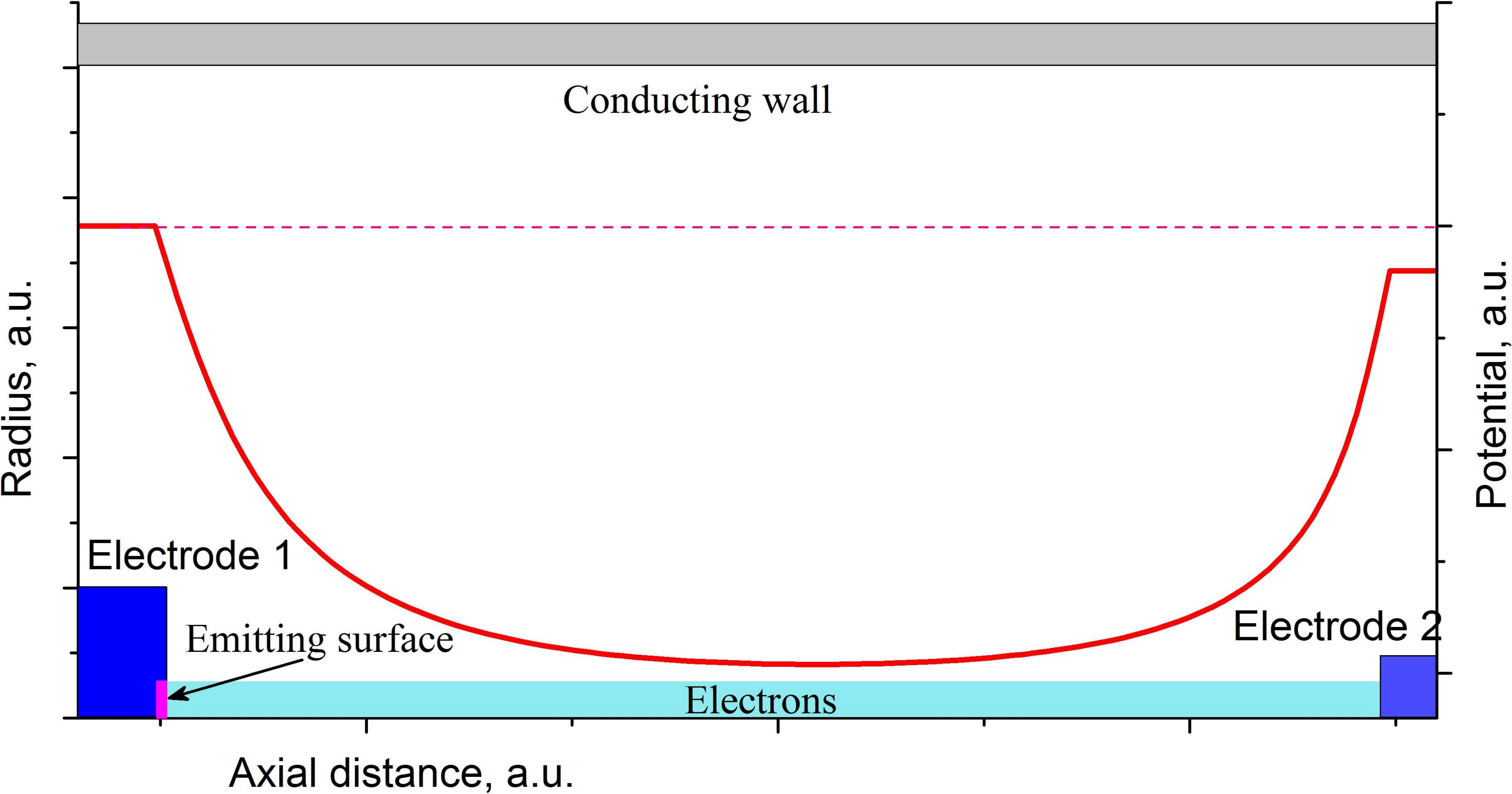}}
	\caption{ Schematic drawing of the system. Two electrodes are placed inside a conducting tube. The potentials of the electrodes are negative relative to that of the tube and are slightly different. The potential distribution along the axis of the tube is shown schematically by the red line. A strong axial magnetic field restricts electron transverse motion in the cloud. }
	\label{Fig-1}
\end{figure}

Let us consider the configuration, formed by a conducting tube of radius $R$ and two electrodes placed on the axis and separated by a distance  $L\gg R$ (Figure~\ref{Fig-1}). The electrodes are at negative potential (not necessarily the same) relative to the potential of the tube. The electrons are injected into the tube from the first electrode(left in Fig.~\ref{Fig-1}) and propagate along a longitudinal magnetic field which is sufficiently strong to prevent radial expansion. The negative potential of the second electrode (right) reflects the electrons completely or partially depending on its potential and radius. Any electron which reaches one of two electrodes is absorbed. Thus, the injected electrons, depending on their energy, either leave the system or are trapped in both the radial and longitudinal direction, and form a negatively charged column of space-charge between the electrodes. Electron motion in the transverse direction can be described by azimuthal rotation in the crossed axial magnetic field and radial electric field of the electron space-charge.  Longitudinal motion is defined by oscillation in the potential well, $\varphi(z)$, between electrodes. Note that the distribution of the potential along the axis is similar to that in a split cathode\cite{Siman-Tov,Leopold}. The presence of a radial electric field in the split cathode does not affect the longitudinal electron motion when the applied axial magnetic field is sufficiently strong.

Let the radius $r_d$ of the electron cloud be small, $r_d\ll R$, and assume that the electron number density is homogeneous in the transverse plane \cite{Driscoll, Dubin-1999}. Then,  the radial dependence of the longitudinal component of the electric field, $E_z$, which governs the axial motion of the electrons, can be neglected and the problem  reduces to one-dimensional electron motion. 

Apart from the external electric field, the electron motion is affected by the self-consistent Coulomb electric field of electrons' space-charge. The presence of the conducting tube strongly modifies the force between two charged particles. In contrast to free space, this force decreases much faster than $\propto d^{-2}$, where $d$ is the distance  between particles. Radial homogeneity of the electron number density allows one to consider the electron cloud as an ensemble of charged discs of radius  $r_d\ll R$  with a homogeneous surface charge density $\sigma$. The condition  $r_d\ll R$ guaranties that the force of the Coulomb interaction between these charged discs is distributed almost homogeneously across the disc' surfaces and can be replaced  by its integrated value and the discs can be considered as rigid. Azimuthal rotation in the crossed radial electric and axial magnetic fields does not violate this assumption.

Let us replace a $dz$-thick cylindrical slice of the electron cloud placed at a distance $z_0$ from the emitting electrode (left in Fig.~\ref{Fig-1}) by a charged disc with surface charge density $\sigma_0=en(z_0)dz$, where $n(z_0)$ is the electron number density. The longitudinal electric field $E_z$ produced by this disc at some point $(z,r)$, can be written as\cite{Smythe}: 
\begin{equation}\label{eq1}
E_z(z,r,\phi)=\frac{4\pi \sigma_0 r_d}{R}\, \sum_n e^{-\nu_n|z-z_0|/R}{\rm sign}(z-z_0)\frac{J_0(\nu_nr/R)J_1(\nu_nr_d/R)}{\nu_nJ_1^2(\nu_n)},	
\end{equation}	
where  $J_{0,1}$ are Bessel functions of zero and first order, respectively, and $\nu_n$ are the zeros of $J_0(\nu_n)=0$. 
One can verify that the dependence of the field $E_z$ on radius $r$ is very weak in the region $r\leq r_d\ll R$ near the axis, so that it is possible to use an average over the disc surface value of the field \cite{Williams, Petersen}:
\begin{eqnarray}
\label{eq2}
E_z(z,r)\Rightarrow \bar{E}_z(z)=\frac{2\pi}{\pi r_d^2}\int rdrd E_z(z,r)\nonumber\\
\bar{E}_z(z)=8\pi \sigma_0\,\sum_n e^{-\nu_n|z-z_0|/R}{\rm sign}(z-z_0)\frac{J_1^2(\nu_nr_d/R)}{\nu_n^2J_1^2(\nu_n)}\equiv 8\pi \sigma_0f(|z-z_0|){\rm sign}(z-z_0).
\end{eqnarray}
Finally, one can  write the longitudinal equation of motion of an electron in the presence of the external electric field $E_{\rm ext}(z)=-\partial\varphi(z)/\partial z$ and the electric field of the space charge as:
\begin{equation}\label{eq3a}
\frac{d^2z}{dt^2}=\frac{e}{m}\left[-\frac{\partial\varphi(z)}{\partial z}+8\pi e\int_0^Ldz_0n(z_0)f(|z-z_0|){\rm sign}(z-z_0)\right]
\end{equation}
where $z$ is the electron axial position and $e>0$ is the elementary charge attributed to electrons.

\begin{figure}[tbh]
	\centering \scalebox{0.1}{\includegraphics{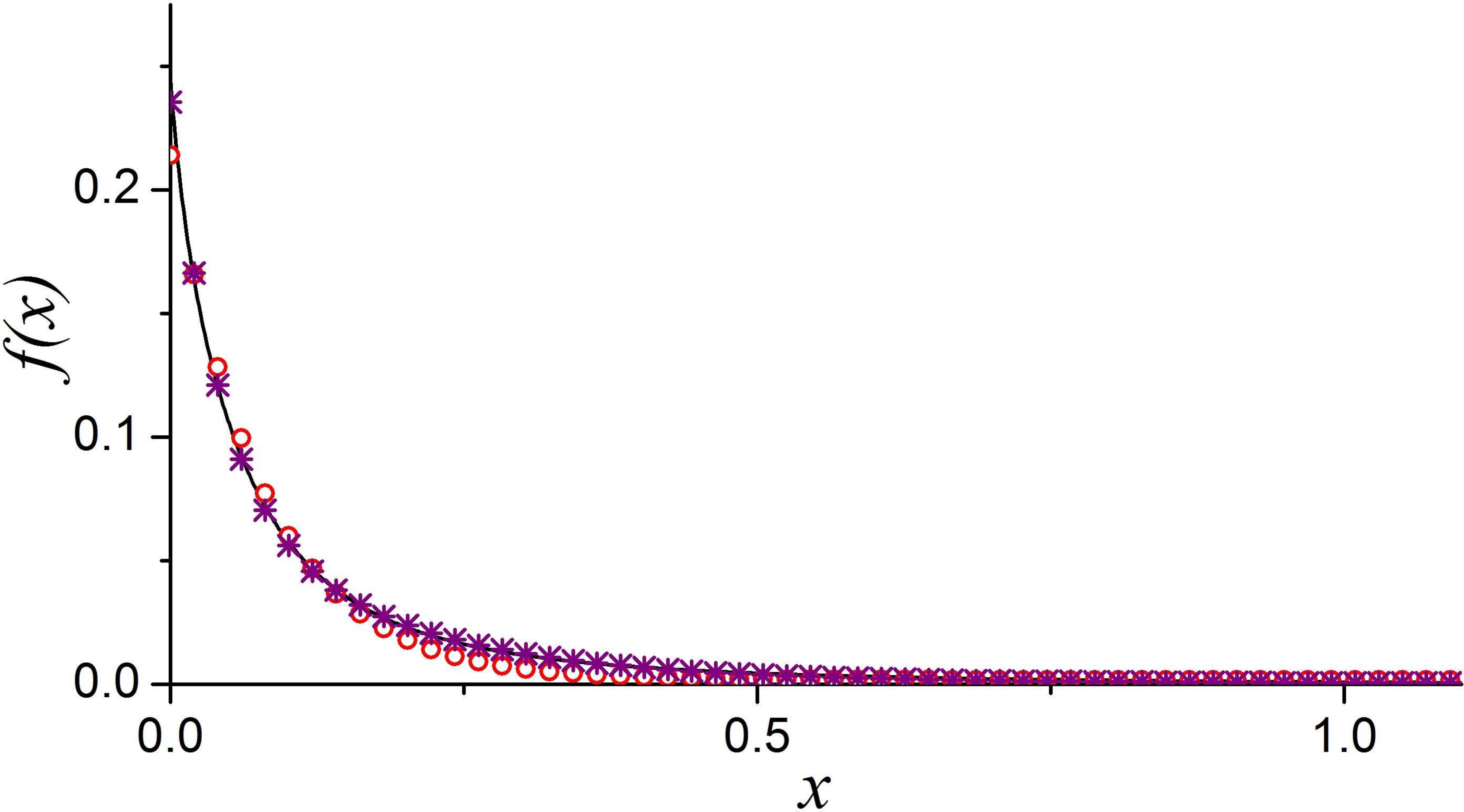}}
	\caption{ The function $f(x)$, $x=|z-z_0|/R$ at $r_d/R=0.1$. Solid line -- calculated using Eq.~(\ref{eq1}); circles -- approximation by one exponent, $f(x)=a\exp(-\lambda x)$; asterisks -- approximation by two exponents, $f(x)=a_1\exp(-\lambda_1 x)+a_2\exp(-\lambda_2 x)$. }
	\label{Fig-2}
\end{figure}

A typical dependence of the function $f(x)$ on the normalized distance  $x=|z-z_0|/R$, is shown in Fig.~\ref{Fig-2}.
Figure~\ref{Fig-2} shows that the longitudinal electric field of the space charge decreases almost exponentially with the distance from a charged slice (disk). Because the effect being studied does not depend on the fine structure of the force, we will use below the exponential approximation
\begin{equation}
\label{eq4a}
f(x)=a\exp(-\lambda x).
\end{equation}
For example, for $r_{d}/R=0.1$, $a\simeq0.25$ and $\lambda\simeq 13$. 

Introducing the dimensionless variables: 
$$\xi=z/R,\,\tau=\frac{t}{R}\sqrt{\frac{2e\varphi_m}{m}},$$ 
where $\varphi_m$ is the depth of the potential well, the equation of motion Eq.~(\ref{eq3a}) takes the following form:
\begin{equation}
\label{eq5b}
\frac{d^2\xi}{d\tau^2}=-\frac{1}{2}\frac{d\Phi}{d\xi}+4\frac{R^2}{r_d^2}\int_0^{\mathcal{L}}\,d\xi_0\,\eta(\xi_0)f(|\xi-\xi_0|){\rm sign}(\xi-\xi_0).
\end{equation}
Here $\Phi=\varphi/\varphi_m$ and  $\eta=(\pi er_d^2/\varphi_m)n$ are the normalized potential and electron number density respectively, and $\mathcal{L}=L/R$.

If the characteristic spatial scale $\lambda^{-1}$ of the Coulomb interaction is small compared to the characteristic scale of the electron density variation (which is the same as the spatial scale of the variation of the potential $\Phi$), and point $\xi$ is placed far from the potential well edges, the integral in Eq.~(\ref{eq5b}) can be evaluated as
\begin{equation}
\label{eq6}
\int_0^{\mathcal L}d\xi_0\,\eta(\xi_0)f(|\xi-\xi_0|){\rm sign}(\xi-\xi_0)\simeq -2a\frac{d\eta(\xi)}{d\xi}\int_0^\infty\,dx\,x\exp(-\lambda x) =-2a\frac{d\eta(\xi)}{d\xi}\frac{1}{\lambda^2}.
\end{equation}
Now Eq.~(\ref{eq5b}) can be written as
\begin{equation}
\label{eq7}
\frac{d^2\xi}{d\tau^2}=-\frac{1}{2}\frac{d\Phi}{d\xi}-8\frac{R^2}{r_d^2}\frac{a}{\lambda^2}\frac{d\eta}{d\xi}.
\end{equation} 
Equation (\ref{eq7}) shows that there is a steady-state distribution of the electron number density $\eta_{\rm sq}(\xi)$  whose space-charge electric field compensates  the external electric field completely:
\begin{equation}
\label{eq8}
\eta_{\rm sq}(\xi)=\eta_0-\frac{1}{16}\frac{r_d^2}{R^2}\frac{\lambda^2}{a}\Phi(\xi).
\end{equation}
Let $\Phi(0)=0$. Then $\eta_0=\eta_{\rm sq}(0)$. Equation (\ref{eq7}) is not valid near the edges of the potential well and the integral at $\xi=0$ can be evaluated in a different manner as
\begin{equation}
\label{eq9a}
\left.\int_0^{\mathcal L}d\xi_0\,\eta(\xi_0)f(|\xi-\xi_0|){\rm sign}(\xi-\xi_0)\right|_{\xi=0}\simeq -a\eta_0\int_0^\infty\,d\xi_0\,\exp(-\lambda \xi_0) =-a\eta_0\frac{1}{\lambda}.
\end{equation} 
Thus, at the steady-state:
\begin{equation}
\label{eq10a}
-\frac{1}{2}\left.\frac{d\Phi}{d\xi}\right|_{\xi=0}-4\frac{R^2}{r_d^2}\frac{a}{\lambda}\eta_0 =0
\end{equation}
and 
\begin{equation}
\label{eq11}
\eta_{\rm sq}(\xi)=\frac{1}{16}\frac{r_d^2}{R^2}\frac{\lambda^2}{a}\left[2\frac{\lambda_\Phi}{\lambda}-\Phi(\xi)\right],
\end{equation} 
where $\lambda_{\Phi}=d\Phi/d\xi|_{\xi=0}\ll\lambda$.
Expression (\ref{eq11}) describes the electron number density distribution of the squeezed state of the electron cloud in a longitudinal potential well. 

The squeezed state forms and develops in the following way. Let the electrons be injected from the left electrode with some initial velocity into the longitudinal potential well bounded by the absorbing electrodes (Fig.~\ref{Fig-1}). A single electron, depending on its energy, can either reach the absorbing boundary or be reflected  into the space between the electrodes. Note that, the time-dependent longitudinal distribution of the electron number density produces a time varying space-charge electric field, so the total energy of the electrons is not conserved. 
If many electrons are simultaneously present in the well, the self-consistent space-charge field can modify their motion significantly. Part of the electrons  increase their energy, overcome the potential barrier near one of the electrodes, and gets absorbed. There are also electrons whose energy decreases and these are trapped in the potential well even if their initial energy exceeds the potential barrier. Accelerated electrons leave the system, while the decelerated electrons accumulate in the potential well. The accumulated electrons decrease the depth of the effective potential well, $\varphi_{\rm eff}$, which includes the contribution of space charge. Finally, the potential well becomes completely filled with low-energy electrons, so that the effective potential  $\varphi_{\rm eff}\simeq {\rm const}$  and $E_{\rm eff}\simeq 0$, a situation which defines a squeezed state.

\section{Numerical simulations}

We solve here the equation of motion Eq.~(\ref{eq5b}) in its discrete form numerically. 
The energy of the injected electrons (macro particles) is chosen to be small relative to the potential well depth. 
The number of injected electrons per unit time is kept constant, which corresponds to  saturated thermionic emission. When the number of electrons accumulated in the potential well is sufficiently large, part of the injected electrons returns back immediately after injection, so that the actual emitted current becomes space-charge limited.
The potential of the right boundary of the potential well can be chosen to be either slightly less than or greater than the initial energy of the injected electrons.

\begin{figure}[tbh]
	\centering {\scalebox{0.25}{\includegraphics{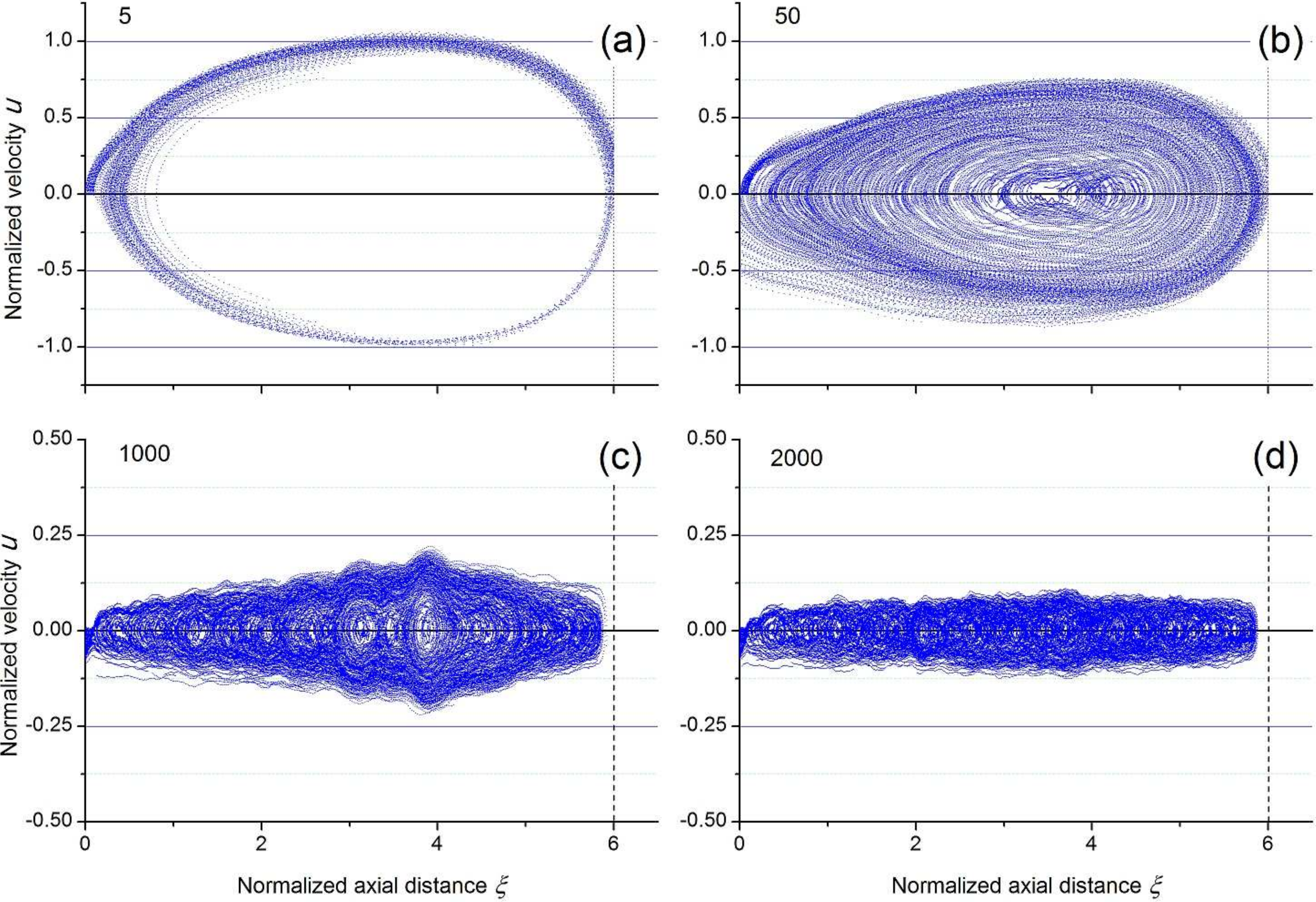}}}
	\caption{ The $(\xi,u)$ phase space at different normalized times $\tau$, (shown at the left top corner of each frame) starting close to the beginning of the injection of electrons. $\tau=5$ in (a), 50 in (b), 1000 in (c) and 2000 in (d). ). [Note the reduced scale in (c) and (d) compared to (a) and (b).]  }
	\label{Fig3a}
\end{figure}

Let us assume that the energy of the injected electrons is chosen to be small relative to both left and right potential barriers or even equal to zero. 
At early times [Fig.~\ref{Fig3a}(a)], electron trajectories in phase space, $(\xi,u=d\xi/d\tau)$, are very near the single electron regular closed trajectory. At the right electrode, $\xi=6$, electrons are either absorbed or return toward the injection point. This behavior keeps on for some time followed by the trajectories filling the phase space area enclosed by the regular trajectory with decreasing energy electrons  [Fig.~\ref{Fig3a}(b)]. Much later, Figs.~\ref{Fig3a}(c) and \ref{Fig3a}(d), the occupied phase space shrinks along the normalized velocity $u$. Note the scale change in Figs.~\ref{Fig3a}(c) and \ref{Fig3a}(d) compared to Figs.~\ref{Fig3a}(a) and \ref{Fig3a}(b). 

As the number of electrons whose energy is large enough to escape from the potential well decreases gradually, the velocity spread $\delta u$ decreases [Fig.~\ref{Fig3a}(c)] until the phase space area diminishes close to the $u=0$ line. Evolution of the phase space is characterized by the formation of numerous vortex-type structures, similar to virtual cathodes,  during the electrons' motion along the potential well, splitting and merging.

\begin{figure}[tbh]
	\centering{ \scalebox{0.2}{\includegraphics{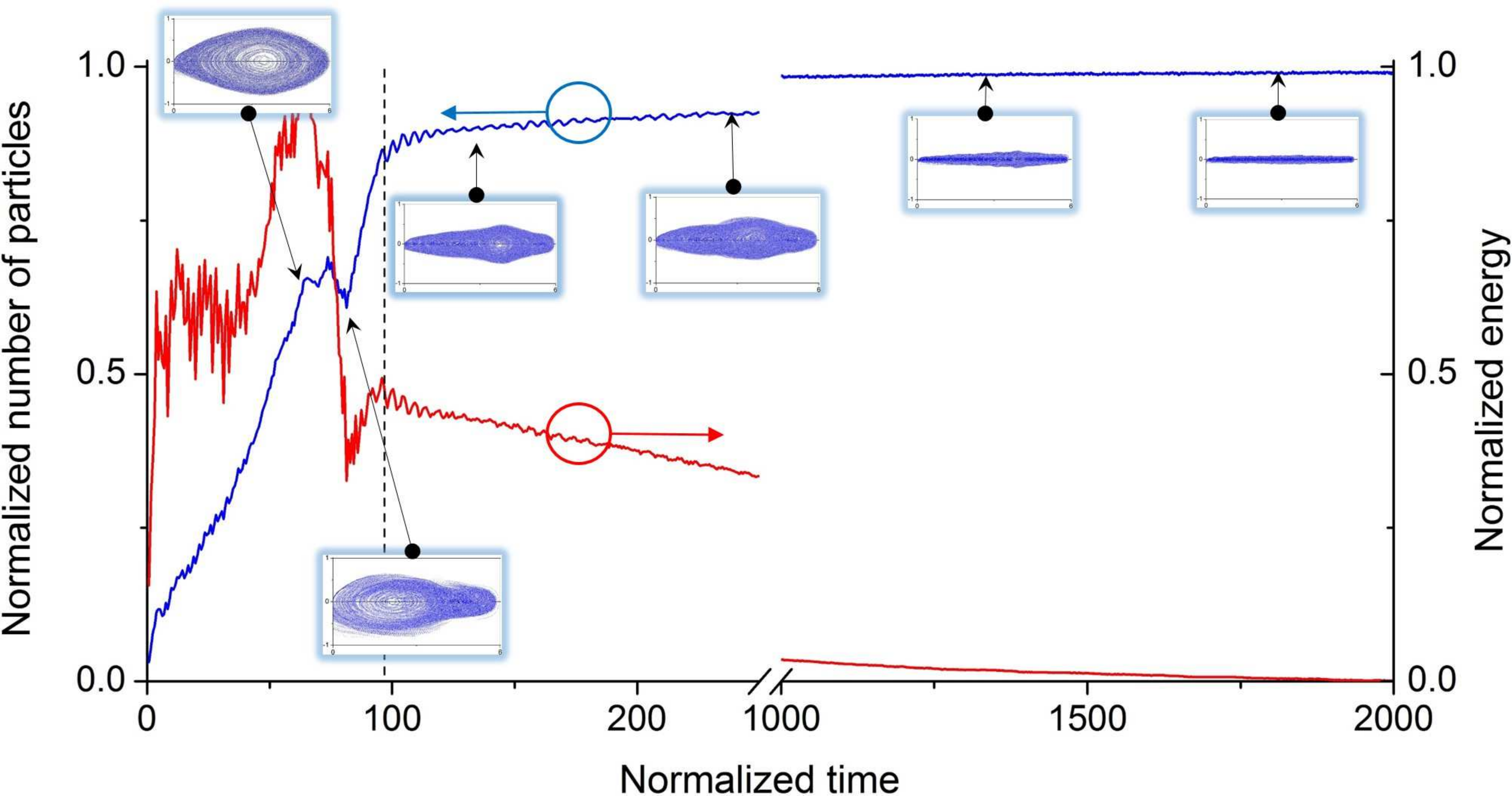}}}
	\caption{The number of electrons (left vertical axis and blue curve) and their total kinetic energy in the potential well (right vertical axis and red curve). Inserts show the phase space evolution at various points in time. The number of particles is normalized to their steady-state value. Time is normalized to the electron transit time along the potential well. The energy is normalized to the potential well depth. The beginning of the squeezed state formation, $\tau_s$, is marked by the dashed vertical line. The normalized velocity axis range in the phase space is [-1,1] for all inserts.}
	\label{Fig.3-2}
\end{figure}

At the initial stage of the potential well's filling with injected electrons, the number of particles in the well, $N_e(\tau)$, and their total kinetic energy,  $W_e(\tau)$, change irregularly. However, when the number of electrons $N_e$ approaches its maximal value, dependencies of the functions $N_e(\tau)$ (the normalized number of electrons) and $W_e(\tau)$ (the normalized total energy) on time change their character sharply (See Fig.~\ref{Fig.3-2}). At the same time, the electron phase space starts to shrink so it is sensible to define this characteristic time, $\tau_s$, marked in Fig.~\ref{Fig.3-2} as the beginning of the squeezed state formation. For long times, the squeezed state  slowly approaches the steady-state value of the number of electrons and the energy decreases, which is a ``cooling'' process of the electron cloud. This behavior does not depend on the initial energy and current of the injected electrons (within a certain range of values) for $\tau>\tau_s$.  But  this is not so for $\tau<\tau_s$. The value of $\tau_s$ depends on the injection characteristics. $\tau_s$ is only several times larger than the time for ``ballistic filling'', $\tau_b$, which is equal to the ratio between the maximum number of the electrons in the potential well and the number of injected electrons per unit time. For example,$\tau_b\simeq 30$ electron transit times for the system studied here, so that $\tau_s\simeq 3\tau_b$.

\begin{figure}[tbh]
	\centering {\scalebox{0.12}{\includegraphics{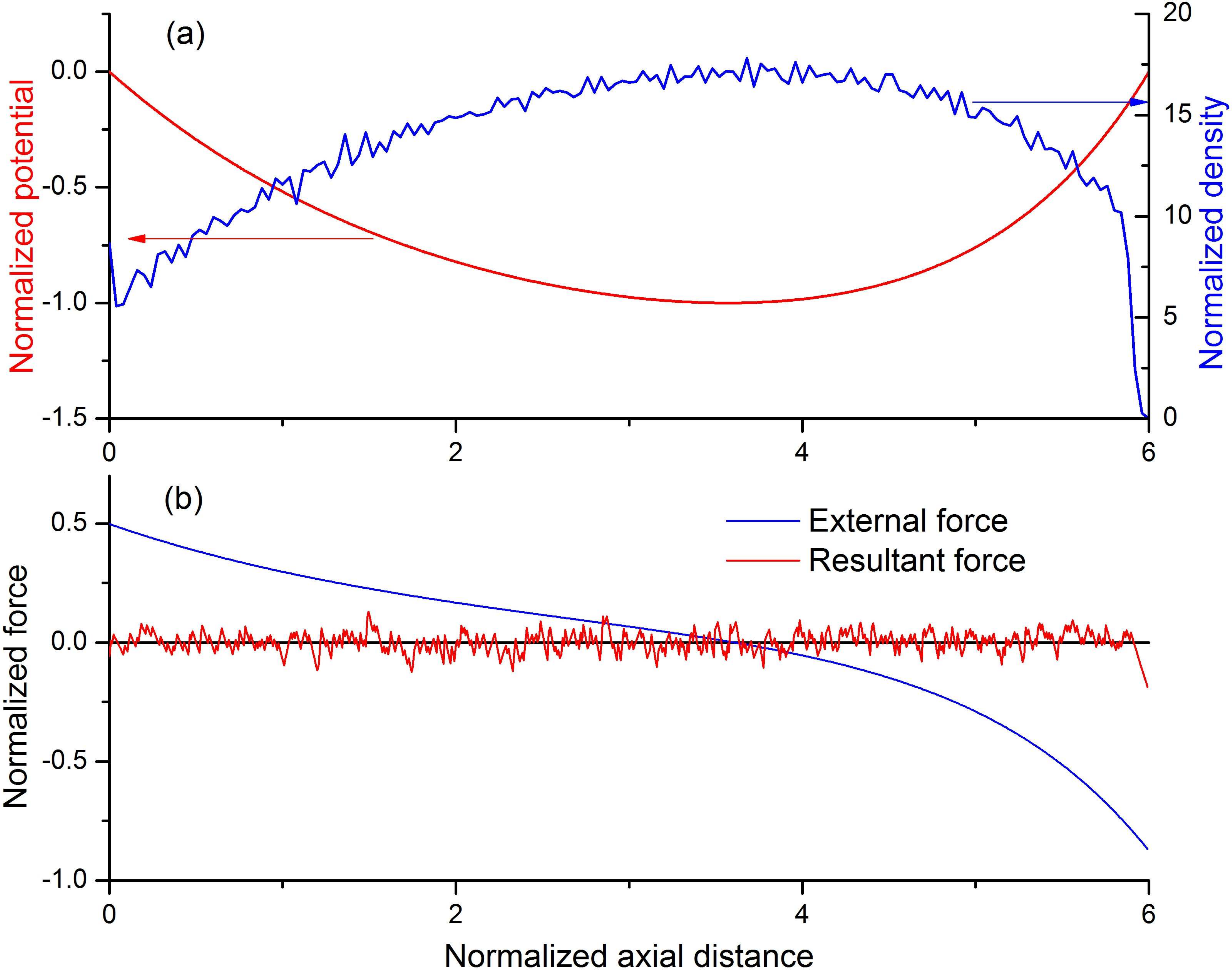}}}
	\caption{(a) The potential well (red) and the electron density  distribution (blue) in the squeezed state. (b)  External (blue) and resultant force (red) inside the potential well. The dimensionless time here is $\tau=20000$. }
	\label{Fig3}
\end{figure}

When the number of trapped electrons is very near the saturation level  (see Fig.~\ref{Fig3a} and the insert at $\tau\simeq 1800$ in Fig.~\ref{Fig.3-2}), the electron number density distribution along the potential well is almost like the mirror image of the external potential [see Fig.~\ref{Fig.3-2}(a)], as predicted by Eq.~(\ref{eq11}). The distribution of the resultant force along the potential well is shown in Fig.~\ref{Fig3}b.

Note, that the maximal number of macro particles present simultaneously in the potential well, is  $\sim 3\cdot 10^3$. This number is sufficient for the correct numerical simulation of the squeezed state formation, including collective effect such as space charge waves (see Section IV). Decreasing the ``weight'' of the macro particles by a factor of two or three, that is, increasing the number of macro particles, did not change the results.

There is an important difference between the electron number density distribution along the potential well in the squeezed state and in the electron column at thermal-equilibrium in the Malmberg-Penning trap. In the squeezed state, the electron density is non-zero at the injection plane, while in the Malmberg-Penning trap the electron column is isolated from the boundaries and the electron density is equal to zero at the boundaries. This is a manifestation of the difference between open and closed systems.

The squeezed state can also form in the absence of the left electrode in Fig. 1. For example, assume that a fixed energy electron beam is injected from the left boundary and is reflected by the right electrode whose potential exceeds the beam energy. This configuration is similar to those considered in Refs.~\cite{Bettega-2007, Dubinov-2020a}. If the beam current is sufficiently large, a virtual cathode forms near the injection plane. This virtual cathode can be considered as a virtual potential barrier near the left boundary. A detailed consideration of such a system is beyond the scope of this article, but preliminary studies carried out in the framework of the one-dimensional model presented above, demonstrate the same phase space evolution as 3D PIC (Particle-in-Cell) simulations\cite{Dubinov-2020a}.

Finally, let us estimate the characteristic electron number density in the squeezed state. Assuming that the tube radius is $R\sim 10$ cm, the electron column radius is $r_d\sim 1$ cm, and the potential well depth $\varphi_m$ is of the order of several tens of kV, one can calculate that the electron number density is of the order of $10^{11}$--$10^{12}$ ${\rm cm}^{-3}$. This electron density is several orders of magnitude larger than one accessible in the Malmberg-Penning trap.

\section{The squeezed state as a one-component  plasma}

The results of Section III confirm that the electric field of the space-charge trapped electrons inside the potential well compensates almost completely the external electric field. One can, instead, consider the formation of the squeezed state as the result of the electric field of the electron space-charge compensated by the external electric field. This is exactly the situation occurring in a two-component quasi-neutral plasma, when electrons and ions compensate their own space-charge and form a force-free configuration with equal densities of the particles. Thus, a squeezed state may be considered as a charged, non-neutral electron plasma.

The electron cloud in the Malmberg-Penning trap was also treated as a ``plasma with a single sign of charge'',\cite{ONeil-1988, ONeil-1999, Dubin-1999} and this analogy was used when considering the radial equilibrium of the electron cloud. In this context, the axial magnetic field plays the role of an ion background compensating the electrons' space-charge. Instead, we use the terminology of a charged plasma by keeping in mind that the axial equilibrium of the electron cloud provides its radial confinement. We should note that some similarity between the squeezed state and plasma was mentioned in Refs.~\cite{Dubinov-2016, Dubinov-2020}.

The interpretation of the squeezed state as a quasi-neutral plasma allows one to assume that some kind of plasma wave can propagate along the electron column. Fig.~\ref{Fig.3-2} shows that the squeezed state relaxes to its steady state very slowly. Intense fluctuations of the electron density and the electric field are very slowly damped and remain significant over a very long time, but these noise-like perturbations make their observation difficult. However, it is possible to arrange the electrons (macro particles) along the potential well in such a way, that the relation between the electric potential and the electron number density [Eq.~(\ref{eq11})] distribution is fulfilled. This artificially prepared initial state relaxes rapidly to a steady-state with very low noise level.

\begin{figure}[tbh]
	\centering\scalebox{0.2}{\includegraphics{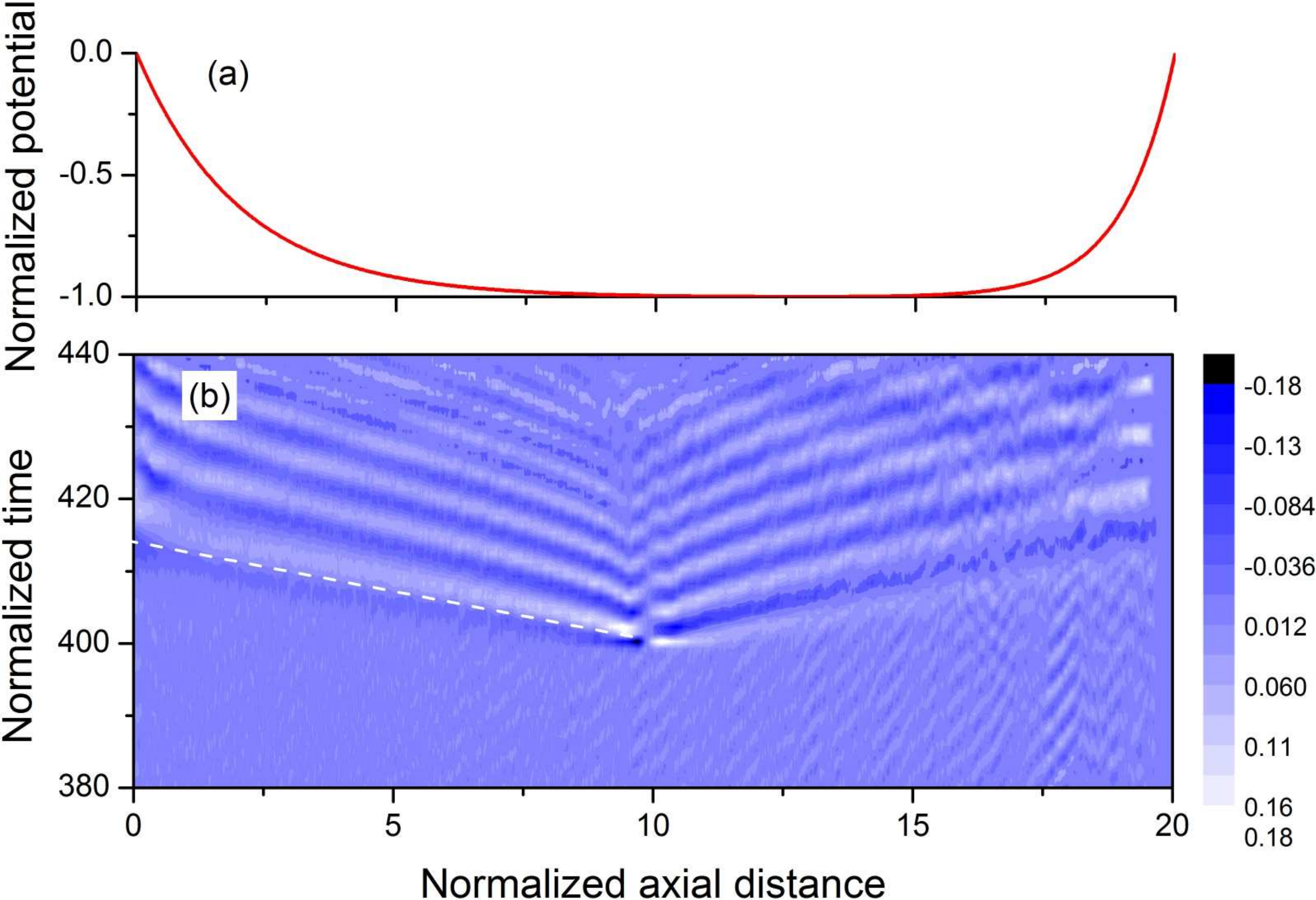}}
	\caption{ (a)  External potential well. (b) The electric field evolution.  Note that the wave phase velocity is constant in the central region, where $\Phi\sim {\rm const}$, and  decreases near the system edges, where the equilibrium density is minimal. The dashed line corresponds to $c_{\rm sq}=1/\sqrt{2}$.}
	\label{Fig9}
\end{figure}

For example, let us consider that the steady state is perturbed by additional particles which appear suddenly in a small region near the center of the potential well as seen in Fig.~\ref{Fig9}. The uncompensated space-charge field produced by these excess particles removes particles from the potential well through the left and right absorbing boundaries and the total number of particles in the squeezed state relaxes rapidly to its equilibrium value. However, the electric field and electron density perturbations remain in the system in the form of propagating waves that are clearly visible in Fig.~\ref{Fig9}(b).

The evolution of small perturbations of the dimensionless steady-state electron density $\tilde{\eta}$ and electron velocity $\tilde{u}$ is described by the equation of motion and the continuity equation:
\begin{eqnarray}
\frac{\partial \tilde{u}}{\partial \tau}=-8\frac{R^2}{r_d^2}\frac{a}{\lambda^2}\frac{\partial\tilde{\eta}}{\partial\xi},\nonumber\\
\frac{\partial \tilde{\eta}}{\partial\tau}+\eta_{\rm sq}(\xi)\frac{\partial\tilde{u}}{\partial\xi}=0,\label{eq9}
\end{eqnarray}
where $\eta_{\rm sq}(\xi)$ is the equilibrium density distribution, defined by Eq.~(\ref{eq11}). 
In the short wavelength approximation, equations (\ref{eq9}) describe waves whose phase velocity, $c_{\rm sq}$,  depends on the local unperturbed density.:
\begin{equation}
\label{eq10}
c_{\rm sq}^2=8\frac{R^2}{r_d^2}\frac{a}{\lambda^2}\eta_{\rm sq}(\xi).
\end{equation}

At the center of the system, where the influence of  boundaries is negligibly small, so that  $\Phi(\xi)\simeq -1$, the phase velocity is constant and it follows from Eq.~(\ref{eq11}) that
\begin{equation}
\label{eq12}
c_{\rm sq}^2\simeq 1/2,
\end{equation}
which is confirmed by the results of numerical simulations shown in Fig.~\ref{Fig9}(b). The dimensional phase velocity, $v_{\rm ph}$, is proportional to the maximum of the electron velocity $v_e=\sqrt{2e\varphi_m/m}$ in the potential well, $c_{\rm sq}=v_e/\sqrt{2}$.    

These waves in the squeezed state are nothing other than the Trivelpiece-Gould modes of a magnetized plasma waveguide \cite{Trivelpiece}.  The maximum phase velocity of these modes, $v_{\rm ph}$, is proportional to the electron plasma frequency $\omega_{pe}\propto n^{1/2}$. In the squeezed state, the electron density, $n$, is proportional to the electrostatic potential $\varphi$. Thus, $v_{\rm ph}\propto v_e \propto\varphi^{1/2}\propto  n^{1/2}\propto \omega_{pe}$. It is not the surprising thing that the electron beam which does not overlap with the squeezed state (a similar configuration with conventional plasma was proposed and used in \cite{Strelkov} ) can excite these modes \cite{Dubinov-2016}.

\section{Conclusions}

A squeezed state can develop when electrons are continuously injected at one or both edges of a longitudinal potential well. The properties of this state of an electron cloud are similar to that of an one-dimensional quasi-neutral plasma where the external electrostatic potential substitutes the role of the ions. Quasi-neutrality of an ordinary, two-component plasma, is replaced by equilibrium between the confining external potential field and the repulsive Coulomb force of the space-charge. It is important to note that our interpretation of the equilibrium condition allows one to determine the maximum number of charged particles which can be accumulated in a given potential trap.

\section*{Acknowledgment}

The research at the Technion was supported by Technion Defense Grant No. 2029541 and ONR Grant No. N62909-21-1-2006.

\section*{Data Availability}

The data that supports the findings of this study are available within the article.


\begin{thebibliography}{99}
\bibitem{Ignatov-1994} A.M. Ignatov and V.P. Tarakanov, Phys. Plasmas {\bf 1}, 741 (1994).	
\bibitem{VC} J. Benford, J.A.Swegle, and E. Schamiloglu, {\it High Power Microwaves}, Tayler \& Francis, (2007). 
\bibitem{Bettega-2007} G. Bettega, F. Cavaliere, M. Cavenago, A. Illiberi, R. Pozzoli, and M. Rom\'e, Phys. Plasmas {\bf 14 }, 042104 (2007).
\bibitem{MP-trap} F.M. Penning,  Physica (Amsterdam) {\bf 3}, 873 (1936).
\bibitem{MP-trap-2} J.S. deGrassie and J.H. Malmberg,  Phys. Rev. Lett. {\bf 39},
1077 (1977).
\bibitem{Shamiloglu} M.I. Fuks, S. Prasad, and E. Schamiloglu, IEEE Trans. Plasma Sci. {\bf 44}, 1298 (2016).
\bibitem{Fuks} M.I. Fuks and E. Schamiloglu, Phys. Rev. Lett. {\bf 122}, 224801 (2019).
\bibitem{Siman-Tov} M. Siman-Tov, J.G. Leopold, Y.P. Bliokh, and Ya.E. Krasik, Phys. Plasmas {\bf 27}, 083103 (2020).
\bibitem{Leopold} J.G. Leopold, M. Siman-Tov, S. Pavlov, V. Goloborodko, Ya.E. Krasik, A. Kuskov, D. Andreev, and E. Schamiloglu, J. Appl. Phys. (accepted for publication) (2021).
\bibitem{Egorov} E.N. Egorov, A.A. Koronovskii, S.A. Kurkin, and A.E. Hramov, Plasma Phys. Rep. {\bf 39}, 925 (2013).
\bibitem{Dubinov-2017} A.E. Dubinov and V P. Tarakanov, Laser Part. Beams {\bf 35}, 362 (2017).
\bibitem{Dubinov-2016} A.E. Dubinov, A.G. Petrik, S.A. Kurkin, N.S. Frolov, A.A. Koronovskii, and A.E. Hramov, Phys. Plasmas {\bf 23}, 042105 (2016).
\bibitem{Dubinov-IEEE} A.E. Dubinov and V.P. Tarakanov, IEEE Trans. Plasma Sci. {\bf 49}, 1135 (2021). 
\bibitem{Dubin-1999} D.H.E. Dubin and T.M. O'Neil, Rev. Mod. Phys. {\bf  71}, 87 (1999).
\bibitem{Driscoll} C.F. Driscoll, J.H. Malmberg, and K.S. Fine, Phys.Rev.Lett. {\bf 60}, 1290 (1988).

\bibitem{Smythe} W.R. Smythe, {\it Static and Dynamic Electricity}, Taylor \& Francis, (1989).
\bibitem{Williams} C.B. Williams and M.H. MacGregor, IEEE Trans. Nuc. Sci. NS-14, 581 (1967).
\bibitem{Petersen} G.W. Petersen and W.J. Gallagher, IEEE Trans. Nuc. Sci. NS-16, 214 (1969).
\bibitem{Dubinov-2020a} A.E. Dubinov and V.P. Tarakanov, Tech. Phys. {\bf 65}, 1002 (2020).
\bibitem{ONeil-1988} T.M. O'Neil, AIP Conference Proceedings {\bf 175}, 1 (1988).
\bibitem{ONeil-1999} T.M. O'Neil, Phys. Today {\bf 52}, 24 (1999).

\bibitem{Dubinov-2020} A.E. Dubinov, V.D. Selimir, and V.P. Tarakanov, Plasma Phys. Rep. {\bf 46}, 1108 (2020).
\bibitem{Trivelpiece} A.W. Trivelpiece and R.W. Gould, J. Appl. Phys. {\bf 30}, 1784 (1959). 
\bibitem{Strelkov}M.V. Kuzelev, F.Kh. Mukhametzyanov, M.S. Rabinovich, A.A. Rukhadze,
P.S. Strelkov, and A.G. Shkvarunets, Sov. Phys. JETP {\bf 56}, 780 (1982). 
	
\end{thebibliography}
\end{document}